\input epsf
\documentclass[twoside]{artikel3}
\usepackage{graphicx}
\usepackage{latexsym}
\usepackage{amssymb}
\usepackage{amsmath}
\usepackage{amsopn}
\usepackage{amscd}
\usepackage{ifthen}
\usepackage{fancyhdr}

 2

\font\headfont=cmss10

\newcommand{\algodate}[3]{ %To be called in users tex file
   \def\year{#1}
   \def\month{#2}
   \def\day{#3}
   \def\optionalday{} % will have a comma afer day if supplied
}

\newcommand{\algoheader}[2]{ % To be called in users tex file
   \lhead[{\textbf{\headfont{#2}}}]{}
   \chead[]{}
   \rhead[]{{\textbf{\headfont{#1}}}}
}

\headwidth=\textwidth
\newlength{\voffsetORIG}
\newlength{\headheightORIG}
\newlength{\topmarginORIG}

\makeatletter      % `@' is now a normal "letter" for LaTeX
\newcommand{\ps@firstpagestyle}{
    \renewcommand{\@oddhead}{}
    \renewcommand{\@evenhead}{}
    \renewcommand{\@evenfoot}{}
}
\makeatother     % `@' is restored as a "non-letter" character

\newcommand{\algotitle}[1]{\def\atitle{#1}}

\newcounter{numauthors}  % Increments by +1, each time macro \algoauthor is 
\newcounter{authnum}
\newcommand{\algoauthor}[3]{
   \setcounter{authnum}{#1}
   \ifthenelse{\value{authnum}=1}
   {
       \def\firstauth{#2}
       \def\firstauthaffil{#3}
   }
   {
   \def\secondauth{#2}
   \def\secondauthaffil{#3}
   }
}

\newcommand{\makeauthoraffils}[1]{
            {\sffamily\bfseries \large\firstauth}
            \par\vspace{-0.5em}
            {\sffamily \firstauthaffil}\par
                \smallskip{\sffamily\bfseries \large\secondauth}
                 \par\vspace{-0.5em}
                {\sffamily \secondauthaffil}\par
}

\newcommand{\algoabstract}[1]{\def\aabstract{#1}}

\newcommand{\algokeywords}[1]{\def\akeywords{#1}}

\newcommand{\algomaketitle}{

 \thispagestyle{firstpagestyle}

 {\sffamily\bfseries \LARGE \atitle}\par\bigskip
 \makeauthoraffils{\value{numauthors}}\bigskip
 {\sffamily\bfseries \month\ \optionalday \year}\par\bigskip
 {\sffamily\bfseries \large Abstract}\par
 \vbox{
    \vbox to 0.1875 in{}
    \vbox{
        \leftskip = 0true in
        \rightskip = 0true in
        \vbox{\noindent\rm{\aabstract}\vfill}
    }
 }
 \ifthenelse{ \not\(\(\equal{\akeywords}{}\)\or\(\equal{\akeywords}{ }\)\) }
 {
     {\sffamily\bfseries \large Keywords: }\rm{\akeywords}
 }
 {}
}

\algodate{2010}{February}{23}

\renewcommand{\theequation}{\thesection.\arabic{equation}}
\newtheorem{theorem}{Theorem}[section]
\newtheorem{lemma}[theorem]{Lemma}

\newtheorem{corollary}[theorem]{Corollary}
\newtheorem{remark}[theorem]{Remark}

\newcommand{\Bid}{{\mathchoice {\rm {1\mskip-4.5mu l}}
{\rm {1\mskip-4.5mu l}} {\rm {1\mskip-3.8mu l}} {\rm {1\mskip-4.3mu l}}}}
\newcommand{\expec}{\ensuremath{\mathbb {E}}}
\newcommand{\var}{\mbox{Var}}
\newcommand{\cO}{{\cal O}}
\newcommand{\G}{\ensuremath{\mathbb {G}}}
\newcommand{\E}{\ensuremath{\mathbb {E}}}
\def\endproof{ \hfill $\Box$}

\begin{document}
\algotitle{Adaptive Simulation of the Heston Model}

\algoauthor{1}{Ian Iscoe}{Quantitative Research, Algorithmics, Inc.}
\algoauthor{2}{Asif Lakhany}{Quantitative Research, Algorithmics, Inc.}

\algoabstract{%
Recent years have seen an increased level of interest in pricing equity options 
under a stochastic volatility model such as the Heston model. Often, simulating 
a Heston model is difficult, as a standard finite difference scheme may lead to 
significant bias in the simulation result. Reducing the bias to an acceptable 
level is not only challenging but computationally demanding. In this paper we 
address this issue by providing an alternative simulation strategy -- one that 
systematically decreases the bias in the simulation. Additionally, our 
methodology is adaptive and achieves the reduction in bias with ``near'' 
minimum computational effort. We illustrate this feature with a numerical 
example.}

\algokeywords{stochastic volatility models, efficient simulation}
%\subclass{65C20 \and 65C30 \and 91G60}

\algomaketitle

\algoheader{Adaptive Simulation}{Ian Iscoe \& Asif Lakhany}

\section{Introduction}
\label{introduction}
Under the standard Black-Scholes framework, the asset price dynamics 
is given by the lognormal model
\begin{equation}
\frac{dS_t}{S_t} = \mu dt + \sigma dW_t
\label{eqn_lognormal}
\end{equation}
where the drift parameter $\mu$ and the volatility $\sigma$ are
considered constant (or at best piecewise constant). The popularity
of this model lies in its convenience and simplicity; however, 
these features come at a price. The standard Black-Scholes model is not able to
capture the volatility smile observed in the trading market. The Heston
model assumes that the variance $V = \sigma^2$ is itself a stochastic process --
more specifically a square-root diffusion model of the CIR (Cox-Ingersoll-Ross)
type (see \cite{Heston}, \cite{CIR}). It further allows the
variance, $V$, to be correlated with the stock price $S$, thereby capturing
the volatility smile. Furthermore, the Heston model provides a closed-form
solution for pricing European Options, allowing one to fit the
model to observed option prices. The Heston model is given by the
coupled SDE (stochastic differential equations):
\begin{eqnarray}
\frac{dS_t}{S_t} &=& \mu dt + \sqrt{V_t} dW_t^S \label{eqn_hestonS}\\
dV_t &=& \kappa(\theta - V_t)dt + \sigma_V \sqrt{V_t} dW_t^V
\label{eqn_hestonV}
\end{eqnarray}
where the variance $V$ is modelled by a square-root diffusion process with
parameters $\kappa$ which is the speed at which the process mean reverts to the 
long term variance $\theta$, and $\sigma_V$, the volatility of the variance.
We denote by $\rho$ the instantaneous correlation between the two noise 
processes: $d\langle W^S,W^V\rangle = \rho\,dt$. 

By Cholesky factorization, one can rewrite equations 
(\ref{eqn_hestonS})--(\ref{eqn_hestonV}) as: 
\begin{eqnarray}
\frac{dS_t}{S_t} &=& \mu dt + \sqrt{V_t}\left[ \rho dW_t^{(1)} + 
\sqrt{1-\rho^2}dW_t^{(2)} \right] \label{eqn_heston_choleski_S}\\
dV_t &=& \kappa(\theta - V_t)dt + \sigma_V \sqrt{V_t} dW_t^{(1)}
\label{eqn_heston_choleski_V}
\end{eqnarray}
with independent Brownian motions $dW_t^{(1)}$ and $dW_t^{(2)}$.
There is a wide range of literature on the simulation of this model. Some
of these are based on Euler discretisation; others on the improved finite 
difference approximations such as higher-order Milstein schemes and 
Predictor-Corrector methods; and still others based on distributionally exact
(\cite{BroadieKaya}) or approximate (\cite{Andersen}) simulation
methods. In our study we focus mainly on the Exact simulation method
as proposed by Broadie and Kaya \cite{BroadieKaya} but we will avoid the 
numerical inversion of the Laplace Transform which seems to be the most time 
consuming part of their simulation technique. Doing so, will mean introducing 
bias in the simulation. We will address this problem by providing an adaptive 
strategy that systematically controls the bias in the simulation. Furthermore, 
this is achieved with minimal computational overhead. In most cases, our method 
should prove faster than that of \cite{BroadieKaya}, as there is no costlier 
inversion to perform. 

Here is the layout of the rest of the paper.
In Section~\ref{sec_stockprice}, we recall the general algorithm 
(cf. Broadie and Kaya \cite{BroadieKaya}) for simulating 
the Heston model, to emphasise the basic difficulty: the simulation of
the integral of the variance process. 
In Section~\ref{sec_bessel} we focus on the dynamics of the variance process and
study its properties by transforming it into a canonical process. 
We also describe methods for simulating the latter process. In 
Section~\ref{sec_besselbridge} we describe a method for simulating bridges 
corresponding to the latter process, and their use in
simulating the integral of the variance process. 
In Section~\ref{sec_theo_integrals} we relate the integral of the variance 
process to a weighted integral of the
canonical process, and we also calculate some moments of the latter for the
canonical bridge process. In Sections~\ref{sec_adaptive}, 
\ref{sec_adapt_efficient}, we explore adaptive computation of the integral. 
In Section~\ref{sec_num_expmnts}, we describe the results of 
some numerical 
experiments on the accuracy and efficiency of our adaptive algorithm.
Finally, we provide concluding remarks and future research ideas in 
Section~\ref{conclusion}. Lengthy or highly technical proofs are given in 
Appendices A--C.

\section{Simulating the Stock Price}\label{sec_stockprice}
\setcounter{equation}{0}
All the details of the simulation algorithm for the equity price
governed by the stochastic differential equation in 
(\ref{eqn_heston_choleski_S}), are given in \cite{BroadieKaya}. We reproduce 
them here for the sake of completeness and to emphasise the role played by the 
integral of the variance process, $V$. Integrating (\ref{eqn_heston_choleski_S})
between two dates of interest (e.g., coupon/reset dates), $t_{j-1}$ and 
$t_j$, we obtain
$$
S_j
 = S_{j-1} \exp\left[ \mu \Delta t_j -\frac{1}{2}\int_{t_{j-1}}^{t_j}\! V_s\,ds
   +\rho \int_{t_{j-1}}^{t_j}\!\sqrt{V_s}\,dW_s^{(1)} 
   + \sqrt{1-\rho^2} \int_{t_{j-1}}^{t_j}\!\sqrt{V_s}\,dW_s^{(2)}\right]
$$
where $\Delta t_j = t_j - t_{j-1}$. 
Similarly, integrating (\ref{eqn_heston_choleski_V}) we obtain
\begin{equation}
V_j
 = V_{j-1} + \kappa\theta\Delta t_j - \kappa \int_{t_{j-1}}^{t_j}\! V_s\, ds 
   + \sigma_V \int_{t_{j-1}}^{t_j}\! \sqrt{V_s}\,dW_s^{(1)}.
\label{eqn_hestonV_integ}
\end{equation}
In the next section we will see that it is not complicated to simulate $V_j$ in
a manner that is not based on (\ref{eqn_hestonV_integ}). 
We can easily use (\ref{eqn_hestonV_integ}) to obtain:
\begin{equation}
\int_{t_{j-1}}^{t_j}\! \sqrt{V_s}\,dW_s^{(1)} 
 = \frac{1}{\sigma_V}
     \left(
       \Delta V_j - \kappa\theta\Delta t_j +\kappa \int_{t_{j-1}}^{t_j}\!V_s\,ds
     \right)
\label{eqn_distr_int_rtV}
\end{equation}
where $\Delta V_j = V_j - V_{j-1}$. The only component left is
$\int_{t_{j-1}}^{t_j}\! \sqrt{V_s}\,dW_s^{(2)}$; but since $V_s$ is independent
of the Brownian increments $dW_s^{(2)}$ by construction, then given 
$V_s, 0\le s \le t_j$,
$$
\int_{t_{j-1}}^{t_j}\! \sqrt{V_s}\,dW_s^{(2)} 
 \sim N\!\left(0, \int_{t_{j-1}}^{t_j}\! V_s\,ds\right).
$$
Using this result we have:
\begin{equation}
S_j = S_{j-1}\exp\left[
    \mu\Delta t_j -\frac{1}{2}\int_{t_{j-1}}^{t_j}\! V_s\, ds+
    \rho \int_{t_{j-1}}^{t_j}\! \sqrt{V_s}\,dW_s^{(1)}+
    \sqrt{\left(1-\rho^2\right)\int_{t_{j-1}}^{t_j}\! V_s\, ds}\, Z_j\right]
\label{eqn_stock_sim}
\end{equation}
where, given $V_s, 0\le s \le t_j$, $Z_j \sim N(0,1)$ and is conditionally 
independent of $W^{(1)}$.  Based on (\ref{eqn_distr_int_rtV}) and 
(\ref{eqn_stock_sim}), it is clear that simulation of the Heston model is 
straightforward except for the simulation of the integral 
$\int_{t_{j-1}}^{t_j}\!V_s\,ds$. In \cite{BroadieKaya}, it is done 
in an unbiased manner by using the distribution of this integral. The 
distribution is not available in closed form and is computationally intensive to
compute. Our approach is to use a random quadrature with an adaptive control of
bias.

\section{Squared Bessel Process and its Simulation}
\label{sec_bessel}
\setcounter{equation}{0}

Our first task is to be able to simulate equation (\ref{eqn_heston_choleski_V})
exactly. In order to do so, we cite the following result from \cite{PY}:
\begin{theorem}\label{thm_pitmanyor1}
Consider the one-dimensional diffusion process with laws 
$\{{}^\beta Q^d_x, x\ge 0\}$ 
($x$ is the initial point; $d\ge 0$ and $\beta$ are fixed parameters,
the former called the dimension), with infinitesimal generator 
$$
2xD^2 + (2\beta x+d)D.
$$
Then for all real $\beta \ne 0$, 
${}^\beta Q^d_x$ is the $Q_x^d$-law of the process,
$$
\exp(2\beta t)X\!\!\left( \frac{1 - \exp(-2\beta t)}{2\beta} \right)
$$
where $Q_x^d$ is the distribution of the $d$-dimensional squared Bessel Process 
denoted by {\it BESQ}$^d$ with infinitesimal generator,
$$
2xD^2 + dD.
$$
\end{theorem}
\begin{corollary}
\label{cor_ianasif1}
The space-time transformation
\begin{eqnarray}
V_t &=& \exp(-\kappa t)X_{\tau(t)}, \qquad V_0=x_0, \label{eqn_transfspace}\\
\tau(t) &=& \frac{\sigma_V^2}{4\kappa}\left[\exp(\kappa t)-1\right],\qquad
              \tau_0 = t_0 = 0 \label{eqn_transftime}
\end{eqnarray}
transforms the square-root diffusion process in (\ref{eqn_heston_choleski_V}) to
the $\lambda$-dimensional squared Bessel process:
\begin{equation}
dX_u = \lambda du + 2\sqrt{X_u} dW_u, \qquad X(0) = x_0=V_0
\label{eqn_besselprocess}
\end{equation}
where $\lambda = \frac{4\kappa\theta}{\sigma_V^2}$.
\end{corollary}
\noindent {\bf Proof\ } A direct proof of this result is deferred to Appendix A.
\endproof

The squared Bessel process in (\ref{eqn_besselprocess}) will also be referred to
as a squared Bessel process of order $\nu \equiv \lambda/2 - 1$. From 
\cite{Borodin} we know that the boundary point $0$ is strongly reflective when 
$-1 < \nu < 0$ and 
is ``entrance-not-exit" when $\nu \ge 0$. Following \cite{Yuan} and \cite{Roman}
we simulate the squared Bessel process over any interval $(\tau_j, \tau_{j+1}]$
with $x(\tau_j)$ known (already simulated for $j>0$),
using the randomized Gamma distribution of the first kind, 
$G(\nu+\eta+1,2\Delta\tau)$, where $\Delta\tau = \tau_{j+1}-\tau_j$ and
$\eta$ is sampled from the Poisson distribution 
$P(\mu), \mu=x(\tau_j)/2\Delta\tau$:
$$
X(\tau_{j+1}) \sim G(\nu +\eta +1, 2\Delta\tau) \hspace{1 cm}\mbox{with} \hspace{1 cm}
\eta \sim P\!\left(\frac{x(\tau_j)}{2\Delta\tau}\right).
$$
Once we obtain the simulated values of $X$ on the set of points 
$\tau_j=\tau(t_j), j=1, 2,\ldots,N$, we obtain the corresponding $V_j$ values 
using the transformation (\ref{eqn_transfspace}).

\section{Simulation of Bessel Bridge Process}\label{sec_besselbridge}
\setcounter{equation}{0}
As already emphasised, the most important and difficult piece of our algorithm 
is the simulation of the integral $\int_{t_j}^{t_{j+1}}\!V(s)\,ds$. We
propose to do this by recursively applying a Bessel bridge 
simulation, to fill in the intermediate points in the interval, 
$[t_j, t_{j+1}]$, for a random quadrature. Methods for selecting the 
intermediate points are described later in the paper. In this section we 
describe the method of generation of a Bessel bridge process, and its 
application to the quadrature.

Suppose that on any generic interval $[\tau_L,\tau_R]$ we need to insert another
point $\tau_M$: $\tau_L<\tau_M<\tau_R$. The corresponding {\it BESQ}$^\lambda$ 
value $x_M$ is simulated using the randomized Gamma distribution of the second 
kind, $\G(\cdot,\cdot)$ (see e.g., \cite{Yuan}, \cite{Roman}):
$$
  X(\tau_M) \sim 
   \G\!\left(
             \nu+\eta_1+2\eta_2+1, \frac{\Delta\tau}{2\Delta\tau_L\Delta\tau_R}
       \right)
$$
where $\Delta\tau = \tau_R-\tau_L, \Delta\tau_L \equiv \tau_M-\tau_L, 
\Delta\tau_R\equiv\tau_R-\tau_M$, and where $\eta_1$ is sampled from a Poisson 
distribution, $P(\cdot)$, and $\eta_2$ is sampled from a Bessel distribution, 
$B(\cdot,\cdot)$:
\begin{eqnarray*}
  \eta_1 
   &\sim& P\!\left(\frac{1}{2\Delta\tau}
                \left[\frac{\Delta\tau_R}{\Delta\tau_L}x_L+
                      \frac{\Delta\tau_L}{\Delta\tau_R}x_R
                \right]
           \right) \\
   \eta_2 &\sim& B\!\left(\nu, \frac{\sqrt{x_L x_R}}{\Delta\tau}\right)\,.
\end{eqnarray*}

\begin{remark} For simulating the squared Bessel process and the Bessel
bridge process we need efficient Poisson and Bessel random variate generators.
For generating Poisson variates we refer to \cite{KempKemp}, \cite{AhrensDieter}
and \cite{Hormann}. For generating Bessel variates we refer to Section 2 in 
\cite{Roman} and especially to the comments appearing in Section 4 in
\cite{Yuan}.
\end{remark}

Once we have generated intermediate values on $K-1$ intermediate 
points\footnote{These points may be chosen directly in the $\tau$-space, or
in the $t$-space and mapped to $\tau$-space using the one-one mapping,
(\ref{eqn_transftime}).}
$\tau_1, \tau_2, ..., \tau_{K-1}$ ($\tau_0$ corresponds to the left endpoint
and $\tau_K$ corresponds to the right endpoint of the 
integration interval, $[t_j, t_{j+1}]$), 
the integral of the Variance process $V$ may be approximated by its 
(conditional) expectations over the subintervals. Explicitly, with 
$s_k = \tau^{-1}(\tau_k)$,
\begin{equation}
 \int_{t_j}^{t_{j+1}}\!V(s)\,ds 
 \approx \sum_{k=1}^K \E\!
        \left[
            \int_{s_{k-1}}^{s_k}\!V(s)\,ds \,\Big|\, 
            X\!\left(\tau_{k-1}\right) = x\!\left(\tau_{k-1}\right),
            X\!\left(\tau_k\right) = x\!\left(\tau_k\right) 
        \right].
\label{eqn_expectation}
\end{equation}
In the next section, 
we develop a closed form formula for the conditional expectations on the
right-hand side of (\ref{eqn_expectation}).
For the sake of brevity, when the time interval is fixed, we sometimes denote 
the conditional expectation operator in (\ref{eqn_expectation}) as 
$\expec_{x_L x_R}$, where $L$ and $R$ signify the ``frozen'' left- and 
right-hand endpoints of the interval, respectively. 

We now have the strategy for simulating $\int_{t_j}^{t_{j+1}}\!V(s)\,ds$, for 
some given endpoints $t_j$ and $t_{j+1}$, except for the choice of $K$ and the
intermediate points. 
These topics are discussed in Section~\ref{sec_adaptive}. % and \ref{sec_unif}.

\section{Theoretical Results for Integrals of $V,X$}\label{sec_theo_integrals}
\setcounter{equation}{0}
In this section, we collect some theoretical results on the integrals of $V$ 
and $X$. The proofs of some of the results are deferred to the Appendices.
Our first result expresses the variance integral in the 
$(V,t)$ space in terms of the one in $(X,\tau)$ space. We work on a general 
interval, $[t_L,t_R]$ which, in application, could be $[t_j, t_{j+1}]$ or any
of its subintervals upon refinement of a partition of $[t_j, t_{j+1}]$. Set 
$$
 \tau_L = \tau(t_L),\ \tau_R = \tau(t_R).
$$
\begin{lemma}\label{lem_iansproof1}
On any interval $[t_L, t_R]$, the integral of the variance $V$ is described 
in terms of the canonical {\it BESQ}$^\lambda$ process $X$ as:
\begin{eqnarray}
\int_{t_L}^{t_R}\!V_t\,dt 
 &=& \int_{\tau_L}^{\tau_R}\!X(u)w_0(u)\,du
      \label{eqn_transformationintegral}\\
w_0(u)&:=&a_0[1+cu]^{-2};\ a_0:=4\sigma_V^{-2}, c:=4\kappa\sigma_V^{-2}.
\label{eqn_transf_integrand_weight}
\end{eqnarray}
\end{lemma}
\noindent {\bf Proof\ }
Let $\phi = \tau^{-1}$ so that $\phi(u) = t$ corresponds to $\tau(t)=u$, where
$\tau$ is given by (\ref{eqn_transftime}). Now, $dt=\phi'(u)\,du$ and
$$
 \phi'(u) = 1/\tau'((\phi(u)) 
 = \left[\frac{\sigma_V^2}{4}e^{\kappa\phi(u)}\right]^{-1}.
$$
From the relation, $u=\tau(\phi(u))$, we have that 
$u=[\sigma_V^2/4\kappa][e^{\kappa\phi(u)}-1]$ and so
$$
 e^{\kappa\phi(u)}=1+\frac{4\kappa}{\sigma_V^2}u,\quad
 \phi'(u)=\frac{4\kappa}{\sigma_V^2}\left[1+\frac{4\kappa}{\sigma_V^2}u\right]^{-1}
$$
Therefore, by (\ref{eqn_transfspace}),
\begin{eqnarray*}
\int_{t_L}^{t_R}\!\!V_t\,dt 
&=& \int_{\tau_L}^{\tau_R}\!\!\exp(-\kappa\phi(u)) X(u)\phi'(u))\,du \\
&=&\int_{\tau_L}^{\tau_R}\!\!X(u)
       \frac{4}{\sigma^2_V}\left[1+\frac{4\kappa}{\sigma_V^2}u\right]^{-2}\,du\\
&\equiv&\int_{\tau_L}^{\tau_R}\!\!X(u)w_0(u)\,du.
\end{eqnarray*}
\endproof

Although the {\it BESQ}$^{\lambda}$ process and corresponding bridge are both 
temporally
translation invariant, the function $w_0$ in (\ref{eqn_transf_integrand_weight})
is not. Therefore, when transforming the $\tau$-space integral to a standard 
interval, ``$[0, \tau]$'', the integrand in 
(\ref{eqn_transformationintegral}) will be modified, resulting in 
(``$\stackrel{\cal D}{=}$'' denoting equality in distribution)
\begin{eqnarray}
 \int_{t_L}^{t_R}\!\!V_t\,dt 
 &=&\int_0^{\tau}\!\!X(u+\tau_L)w(u)\,du \ \stackrel{\cal D}{=}\ 
 \int_0^{\tau}\!\!X(u)w(u)\,du,\ \tau=\tau_R-\tau_L,
     \label{eqn_transf_integral_canonical}\\
 w(u)&:=&a_1[b_1+c_1u]^{-2};\ 
         a_1 := 4\sigma_V^{-2},\ b_1:=1+4\kappa\sigma_V^{-2}\tau_L,\  
         c_1:=4\kappa\sigma_V^{-2}, \label{eqn_weight_canonical}
\end{eqnarray}
where we have suppressed the dependence on $L,R$ in the notation, $b_1$ and 
$\tau$. It is implicit, for the rightmost integral in 
(\ref{eqn_transf_integral_canonical}), that the end values of $X$
(at times 0 and $\tau$) are the original values, 
shifted from times $\tau_L$ and $\tau_R$, respectively. 

In the implementation of stopping criteria, the conditional variance of the 
integral in (\ref{eqn_transf_integral_canonical}) is required, conditional on
the endpoint values of the bridge. For that, we state a result on the first two 
moments of the integral of the weighted, squared Bessel bridge process.
The proof of this theorem is lengthy and hence deferred to Appendix B.

\begin{theorem}\label{thm_iantheorem2}
For the {\it BESQ}$^\lambda$ process $X$ of order $\nu$, 
frozen on endpoints, $X(0) = x$ and $X(\tau)= y$ (where $\tau>0$ is arbitrary),
and $w$ as in (\ref{eqn_weight_canonical}), we have
\begin{eqnarray}
\lefteqn{
\E_x\!\left[ \int_0^\tau\!X(u)\,w(u)du\,|\,X(\tau)=y\right]
} \nonumber \\ 
&& = (A_1+B_1)\left(\nu+1+\frac{z}{\tau}R_{\nu}\!\left(\frac{z}{\tau}\right)\right) 
  - \frac{(B_1+C_1)x+(2A_1+B_1)y}{2\tau}
\label{eqn_1stMoment}
\end{eqnarray}
and
\begin{eqnarray}
 \lefteqn{\E_x\!
 \left[ \left(\int_0^\tau\!X(u)w(u)\,du\right)^2 \,|\,X(\tau)=y\right]}
  \nonumber\\
&=&
      2[A_1^2+B_1^2+A_1B_1-A_2-B_2] 
     +(A_1^2+B_1^2+2(A_1+B_1)^2-2A_2-2B_2)\nu 
     + (A_1+B_1)^2\nu^2 
        \nonumber\\ 
& &\ 
   +\ \frac{[(B_1+C_1)x+(2A_1+B_1)y]^2}{4\tau^2}
   + \frac{(B_2+C_2-B_1^2)x - (3A_1^2+B_1^2+2A_1B_1-2A_2-B_2)y}{\tau}
       \nonumber\\
& &\    
     +\ (A_1+B_1)^2\frac{z^2}{\tau^2}
     + [(A_1+B_1)^2+2(A_1^2+B_1^2+A_1B_1-A_2-B_2)]\frac{z}{\tau}R_{\nu}\!
     \left(\frac{z}{\tau}\right)
        \nonumber\\ 
& &\ 
     -\  \frac{(B_1+C_1)x+(2A_1+B_1)y}{\tau}(A_1+B_1)
         \left(\nu+1+\frac{z}{\tau}R_{\nu}\!\left(\frac{z}{\tau}\right)\right)
     \label{eqn_2ndMoment}
\end{eqnarray}
where $z=\sqrt{xy}$, $R_\nu(r)$ is the Bessel quotient 
$I_{\nu+1}(r)/I_\nu(r)$, where $I_{\nu}$ is the modified Bessel function of the 
first kind (see \cite{AS}), and the constants $A_i, B_i, C_i$ , $i=1,2$, are 
described in Proposition \ref{prop_integralsABC}, in Appendix B.
\end{theorem}

\begin{remark} It is not evident that the right-hand sides of 
(\ref{eqn_1stMoment}) and (\ref{eqn_2ndMoment}) tend to zero as $\tau$ tends 
to zero.  However, they do; e.g., the first moment tends to zero at a linear 
rate and the variance tends to zero at a quadratic rate. To see this explicitly,
we reformulate the moments in terms of the parameters $A,b,c$ 
(see Proposition~\ref{prop_integralsABC} and Corollary~\ref{cor_Laplace} in 
Appendix B) for which
$A\to 0$ quadratically fast and $c\to 0$ at a linear rate, as $\tau\to 0$.
Then, in the expression for $A_i,B_i,C_i$ ($i=1,2$) we expand,
$$
 \log(b+c) = \log b + \log(1+c/b) 
   = \log b + \frac{c}{b} -\frac{c^2}{2b^2} + \frac{c^3}{3b^3} + \cO(\tau^4).
$$
Writing $c=c_1\tau$, it is then straightforward to check that
\begin{eqnarray*}
 \lim_{\tau\downarrow 0}\frac{A_1}{\tau^2} &=& -\frac{c_1}{b^2} \\
    %\label{eqn_A1_tau}\\
 \lim_{\tau\downarrow 0}\frac{B_1}{\tau^2} &=& \frac{4c_1}{3b^2} \\
    %\label{eqn_B1_tau}\\
 \lim_{\tau\downarrow 0}\frac{C_1}{\tau^2} &=& -\frac{2c_1}{b^2} \\
    %\label{eqn_C1_tau}\\
 \lim_{\tau\downarrow 0}\frac{A_2}{\tau^4} &=& \frac{5c_1^2}{6b^4}\\%\nonumber\\
 \lim_{\tau\downarrow 0}\frac{B_2}{\tau^4} &=& \frac{8c_1^2}{15b^4}\\%\nonumber\\
 \lim_{\tau\downarrow 0}\frac{C_2}{\tau^4} &=& \frac{4c_1^2}{3b^4}\,.%\nonumber
\end{eqnarray*}
\end{remark}

\section{Adaptive Estimate of the Integral 
$\int_{t_j}^{t_{j+1}}\!V(s)\,ds$}
\label{sec_adaptive}
\setcounter{equation}{0}
We return now to the problem stated at the end of 
Section~\ref{sec_besselbridge}; namely, the selection of the intermediate 
(quadrature) points for the estimation of the integral, 
$\int_{t_j}^{t_{j+1}}\!V(s)\,ds$.

There are three aspects to the selection of 
intermediate points, at any stage in the recursion, in an adaptive fashion: 
(i) the manner of refinement (i.e., the geometric placement of an inserted
partition point or points between those already generated); (ii) a choice
of stopping criterion, to decide if a subinterval needs to be further refined; 
(iii) a decision to apply the stopping criteria locally or globally. 
By definition, a {\it local} decision means that a subinterval 
will be refined if the stopping criterion is not met on that interval, whereas a
{\it global} decision means that all intervals will be refined if any of them do
not meet the stopping criterion. 

Regarding (i), we do not force any particular refinement scheme, although for 
the numerical experiments in Section~\ref{sec_num_expmnts}, we use bisection in
$t$-space.
Regarding (iii), 
we restrict our attention to the class of 
locally adaptive schemes, which we denote by {\it ADAPT}. 
The main purpose of the 
current section is to introduce the stopping criteria of aspect (ii).

In general, a stopping criterion on an interval, involves a tolerance, $\delta$,
and a quantity to monitor. The refinement continues as long as the 
quantity being monitored is not within the tolerance. 
In most cases of interest, the tolerance depends on the interval being 
considered; in fact, it may depend on the entire history of refinement that
led to that interval. To clarify the tolerance's dependence on the interval, we 
make a brief digression on the aspect of tolerance in a refinement scheme. 

The initial tolerance, $\delta_0$, for the
``root'' interval $[\tau_j, \tau_{j+1}]\equiv[\tau(t_j), \tau(t_{j+1})]$ will
be user-given.
For the sake of simplicity we shall relabel this interval as $[\tau_0, \tau_1]$
During the refining of $[\tau_0, \tau_1]$, $\delta_0$ is apportioned among the
subintervals created by the refinement, leading to the $\delta$ for each such
subinterval. We next describe one possible and simple set of rules of 
apportionment; other rules are certainly permitted. In particular,
a more complex and efficient set of rules is described in 
Section~\ref{sec_adapt_efficient}.

Denote the current subinterval being monitored, by 
$[\tau_L, \tau_R] \equiv [\tau(t_L), \tau(t_R)]$. If the stopping criterion 
fails on this interval, then it is partitioned into precisely two 
subintervals.  
The typical case is where refinement is by bisection (either in $t$- or 
$\tau$-space) and the apportionment of tolerance is into equal parts.
For example, if the bisection occurs in $\tau$-space, then the interval 
$[\tau_L,\tau_R]$ inherits the tolerance, 
$\delta_0(\tau_R-\tau_L)/(\tau_1-\tau_0)$. Similarly, if the bisection
is carried out in $t$-space, the tolerance is given by 
$\delta_0(t_R-t_L)/(t_1-t_0)$.

The stopping criterion for our
adaptive scheme, hereafter referred to as $ADAPT$, 
is based on the computation of the variance of the integral. If the 
variance of the integral over any interval, as computed from 
(\ref{eqn_transf_integral_canonical}), (\ref{eqn_1stMoment}) and
(\ref{eqn_2ndMoment}), is below the tolerance, $\delta>0$,
available for the interval, we refrain from further partitioning of that 
interval: for the interval, $[\tau_L,\tau_R]$,
\begin{eqnarray}
\lefteqn{\mbox{\it ADAPT$$ stopping criterion}:} \nonumber\\
&&\hspace*{-0.1in}
  \var\!
   \left[
        \int_{t_L}^{t_R}\!V(s)\,du\,|\, X_{\tau_L}=x_L, X_{\tau_R}=x_R
   \right] %\nonumber \\
 \equiv\ 
  \var\!
   \left[ 
      \int_{\tau_L}^{\tau_R}\!X(u)w_0(u)\,du\,|\, X_{\tau_L}=x_L, X_{\tau_R}=x_R
   \right] 
 < \delta.\nonumber \\         
&&
\label{eqn_adaptivitycriterion2} 
\end{eqnarray}  
This criterion is applied to each subinterval $\left[\tau_L, \tau_R\right]$ 
and, if met, no further partitioning of that interval is done. If the criterion 
is not met, then the interval is partitioned.

Alternatively, when the criterion is not met, the interval $[t_L,t_R]$ can be
partitioned; then the resulting subintervals can be mapped to the corresponding 
subintervals of $\left[\tau_L, \tau_R\right]$, for the generation of 
intermediate $X$ values.  

\section{A more efficient adaptive scheme}\label{sec_adapt_efficient}
\setcounter{equation}{0}
In the previous section, we introduced some locally adaptive schemes. These 
schemes are robust in the sense that they recursively refine until the 
monitored quantity is below the relevant tolerance. The quantity that is 
monitored is subadditive in the sense that once the adaptivity algorithm has 
terminated, the sum of the contributions over all the subintervals 
will be smaller than the initial user-given tolerance, $\delta_0$. However, we 
did not address the issue that this total over all the elements may be 
significantly lower than $\delta_0$. In other words, we did not care about the 
minimality of our refinement. Due to the inherent randomness in the quantity 
monitored and thus the placement of the intermediate points, it may not be 
possible to obtain a minimal grid that barely satisfies the adaptivity 
criterion (\ref{eqn_adaptivitycriterion2}), for example.

However, one can do slightly better than the plain adaptivity schemes described 
in the previous section. One can approach, what we choose to describe as a 
{\it near-minimal} grid by introducing a global reservoir of tolerance, 
$\delta_{\cal R}$.
In essence, we allow the cross-subordination of the tolerance assigned to 
various intermediate elements. When an element passes the adaptivity criteria, 
it releases the excess tolerance that it has over the monitored quantity, to the
reservoir. In the testing of subsequent elements, the monitored quantity is
compared against a more lenient tolerance -- one that is the sum of the 
inherited tolerance level, $\delta$ (as usual), plus the reservoir tolerance, 
$\delta_{\cal R}$.

Here is the precise scheme, in algorithmic form.
Suppose we need to evaluate the integral of the 
Variance process on an interval $[t_0, t_1]$ and further suppose the tolerance 
level is set to $\delta_0$. Our algorithm consists of the following steps:
\begin{description}%[Step 6]
 \item[Step 1:] (Initialisation) Create a {\it reserve} tolerance, 
  $\delta_{\cal R}$, (whose role will be explained in Steps 3, 4) and 
  initialise it to zero ($\delta_{\cal R} = 0$). 
  Prepare an empty stack of 5-tuples $(T_L, T_R, X_L, X_R, \delta)$
  and push the initial data, $(t_0, t_1, X(\tau(t_0)), X(\tau(t_1)), \delta_0)$,
  onto the stack, where $X$ is the {\it BESQ}$^{\lambda}$ process.
 \item[Step 2:] If the stack is empty go to Step 6.
 \item[Step 3:] Pop the top element from the stack. Extract the values 
  and check if $\Delta < 0$, where 
  $$
    \Delta := 
      \var\!
      \left[
           \int_{T_L}^{T_R}\!V(s)\,ds \,|\, X(\tau(T_L))=X_L, X(\tau(T_R))=X_R
      \right]
      - (\delta + \delta_{\cal R}) 
  $$
  If satisfied, go to Step 4; otherwise go to Step 5. (Notice the way in which 
  we use the reserve tolerance $\delta_{\cal R}$ to facilitate passing of the 
  stopping criterion for other ``more needy'' elements).
 \item[Step 4:] Set $\delta_{\cal R} = -\Delta$. Go to Step 2.
 \item[Step 5:] Compute the midpoint $T_M = (T_L + T_R)/2$. Sample 
  $X_M=X(\tau(T_M))$ from the {\it BESQ}$^\lambda$ process. Push the elements 
  $(T_L, T_M, X_L, X_M, \delta/2)$ and $(T_M, T_R, X_M, X_R, \delta/2)$ onto 
  the stack and go to Step 3. (Notice the way in which we distribute the 
  tolerance in equal parts to the newly spawned intervals).
 \item[Step 6:] Exit.
\end{description}

As can be seen from this algorithm, we have incorporated the following features 
that allow us to claim that the algorithm will result in a ``near minimum'' 
number of degrees of freedom necessary: 
\begin{itemize}
\item Our strategy is adaptive. An interval only gets subdivided if it fails the
 test; in the event of a failure, it gives out its acquired delta (from its 
 parent interval) to the newly spawned intervals. Thus there is no loss (or 
 waste) of $\delta$.
\item In the event the test is successful on any interval, it only consumes 
 the amount of acquired delta that is necessary to pass the test, and  
 releases the excess to the tolerance reservoir. The accumulated tolerance in 
 the reservoir can be used later to help pass the test on the remaining 
 intervals.
\end{itemize}
Due to the inherent randomness in the above strategy, an adaptive algorithm that
passes under an absolute minimum number of degrees of freedom, may be difficult 
to find.

\section{Numerical Experiments}\label{sec_num_expmnts}
\setcounter{equation}{0}
In this section we support the ideas introduced in this paper by a series
of numerical experiments. To compare the results of our method 
with other methods available, such as the finite difference method, we 
choose the same test problem as the one appearing in \cite{BroadieKaya}
(Table I, Section 4). 
For ease of reference, the input parameter values are: $S=100$, $K=100$,
$V_0=0.010201$, $\kappa=6.21$, $\theta = 0.019$, $\sigma_V = 0.61$, $\rho=-0.7$,
$r=3.19\%$, $T=1.0$ year. Also, as a benchmark, the true option price = 6.8061.

In our tests we compare our methodology with a variant
of the finite difference method which uses a predictor-corrector step for better
convergence. Also, we have chosen to carry out the refinement scheme in the 
$t$-space and the refinement involves simple bisection of the interval in 
question. 

Since our method is based on discretisation, a bias in the numerical 
method is expected. However, in contrast to the other methods that we know of,
our method allows a systematic control of this bias by means of relating the 
bias to another numerically observable quantity: 
the variance of the integral of the variance 
process over the interval in question. Furthermore, we put great effort 
in ensuring that the computational cost of achieving the limit on the 
variance of the integral (and therefore indirectly on the bias in our method)
is kept to a near minimum as explained in Section~\ref{sec_adapt_efficient}. 
To this end, it is clear
that the very first set of tests should demonstrate the ability of our algorithm
to consume as little computational resources as possible. We do this in 
Tables~\ref{table_num_intervals} and \ref{table_unused_tol}. 
For our test problem, we note that the dimension 
of the BESQ process is given by $\lambda = 1.2684$ and the order is 
$\nu = -0.3658$.
The origin is accessible 
and is also reflective. The moderately high value of the volatility of the 
Variance may also result in many paths taking very high values. Based on 
this observation we decided to capture the properties of our algorithm by 
banding the endpoint value. As expected, we see from 
Table~\ref{table_num_intervals}, that 
for a path that starts at a moderate value and reaches a fairly high value,
the number of the intermediate points that need to be inserted, to reduce 
the variance of the integral below a given tolerance, is also very high.
It appears that the number of intermediate points is proportional to the 
absolute difference between the left and the right endpoint values. We also
observe that the distribution of the number of intermediate points is 
tighter for lower endpoint values compared to higher endpoint values.

In Table~\ref{table_unused_tol}, we demonstrate the effect of cross 
subordination of tolerance which forms an essential part of our algorithm.
What we show in Table~\ref{table_unused_tol}, is the tolerance level that 
was wasted by the algorithm; i.e., for the interval in question, the difference 
between the given tolerance by the user, and the sum of the actual variances of 
the integrals over the subintervals. We again study this in terms of bands of 
endpoint values. As to be expected, the wasted tolerance decreases significantly
when more and more intermediate points are inserted (as can be seen with the 
distribution's very short left tail, for high endpoint values). This effect is 
clear because, when inserting more points, more iterations are made and hence
more use is made of the reserve tolerance. This is an important feature 
of our algorithm as it shows that the wastage is minimal when there is 
highest demand for refinement.

\begin{figure}[hb]
\includegraphics[width=0.9\textwidth]{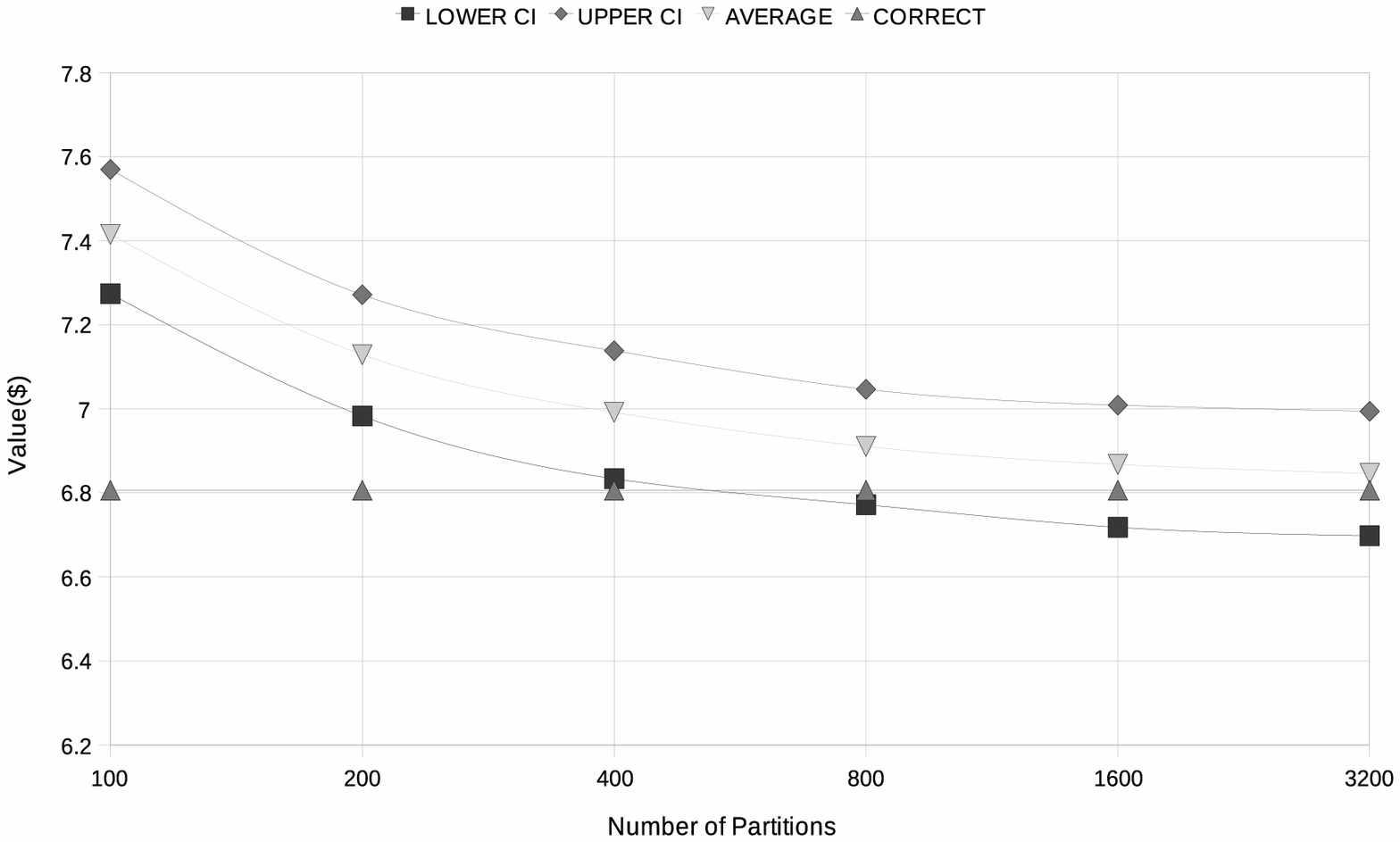}
\caption{Estimation of bias in the predictor-corrector method (1000 trials of 
 10000 samples).}
\label{fig_bias_pred_correct}

\includegraphics[width=0.95\textwidth, totalheight=0.5\textheight]{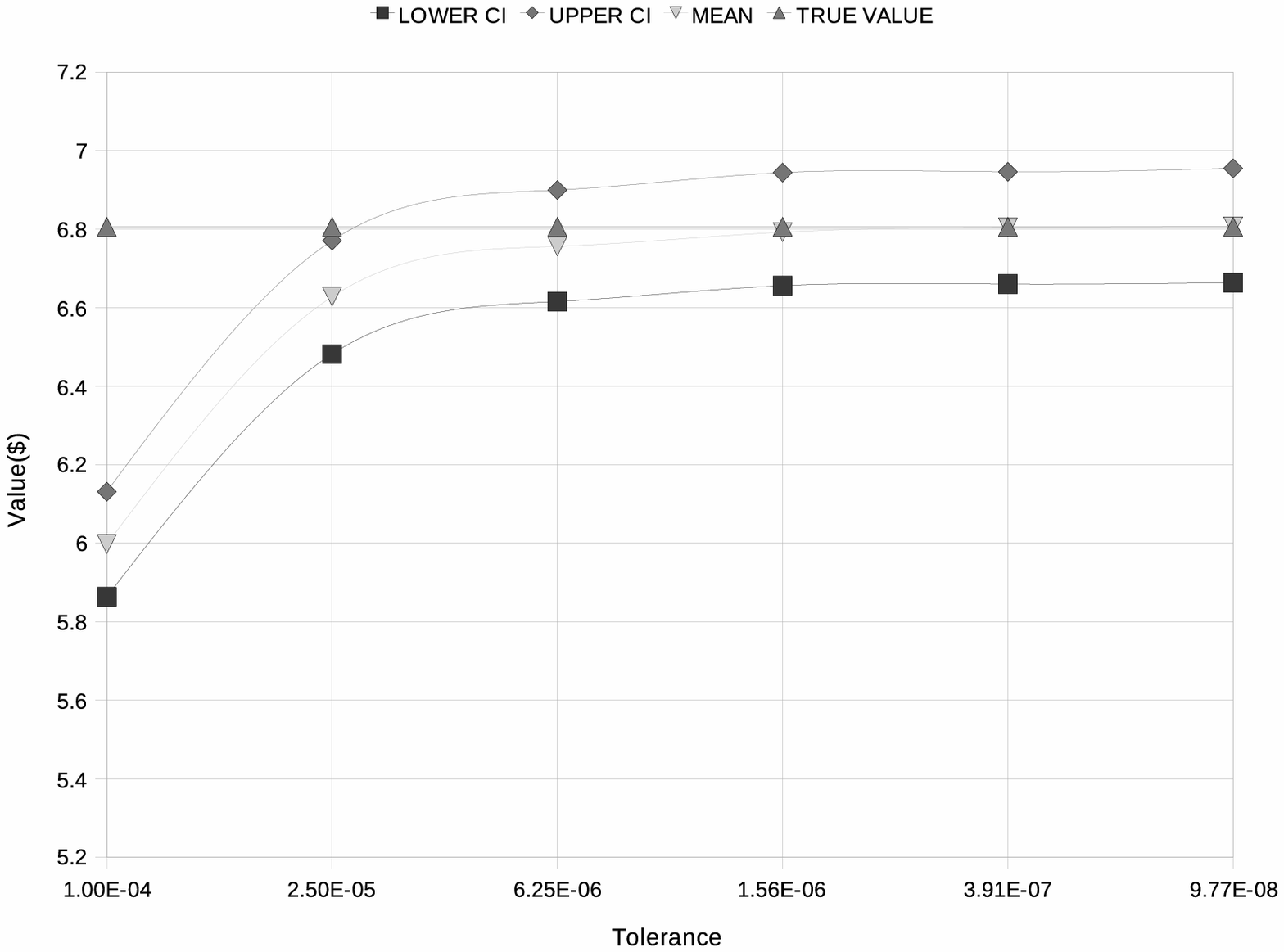}
\caption{Estimation of bias in the ADAPT method (1000 trials of 10000 samples).}
\label{fig_bias_adapt}
\end{figure}

Figures \ref{fig_bias_pred_correct} and \ref{fig_bias_adapt} illustrate the 
expected fact that the bias is reduced in both the predictor-corrector 
method and our adaptive algorithm, as more and more intermediate points are 
introduced. In the predictor-corrector method the independent variable 
is directly the number of intermediate partitions, whereas in our 
algorithm, the independent variable is the tolerance provided by the 
user. Due to this mismatch in the independent variable it is not obvious how
to compare the relative performance of these methods. So in this sense,
Figures \ref{fig_bias_pred_correct} and \ref{fig_bias_adapt} may be considered 
just a sanity check of the expected way in which these algorithms are supposed 
to work. 

\begin{figure}[p]
\begin{minipage}[c]{\linewidth}
\includegraphics[width=0.9\textwidth, totalheight=0.5\textheight]{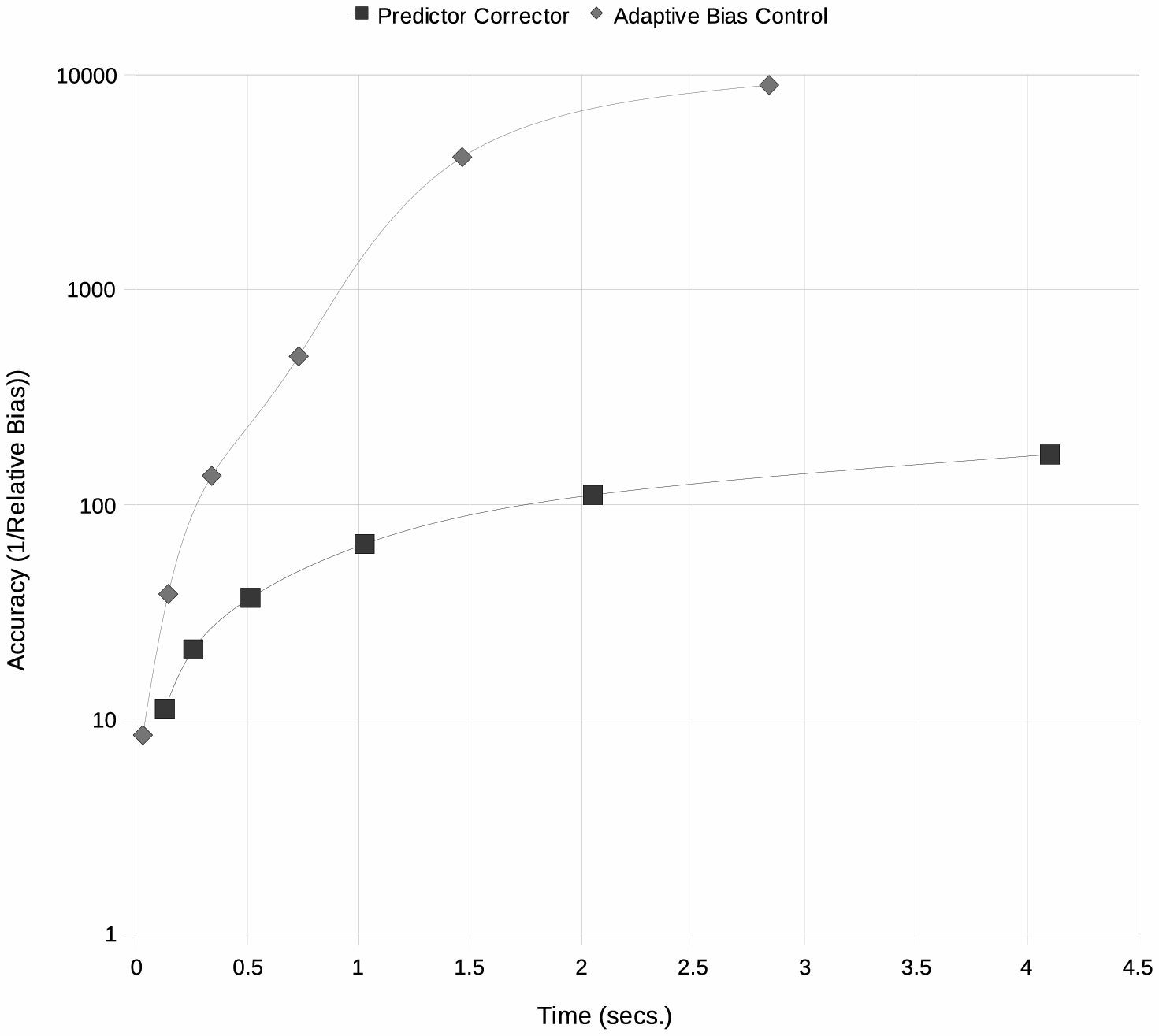}
\caption{Comparison of the predictor-corrector against ADAPT algorithms: 
 accuracy versus time; logarithmic scale on vertical axis.}
\label{fig_pred_correct_vs_adapt}

\vspace*{0.2in}

\includegraphics[width=0.9\textwidth]{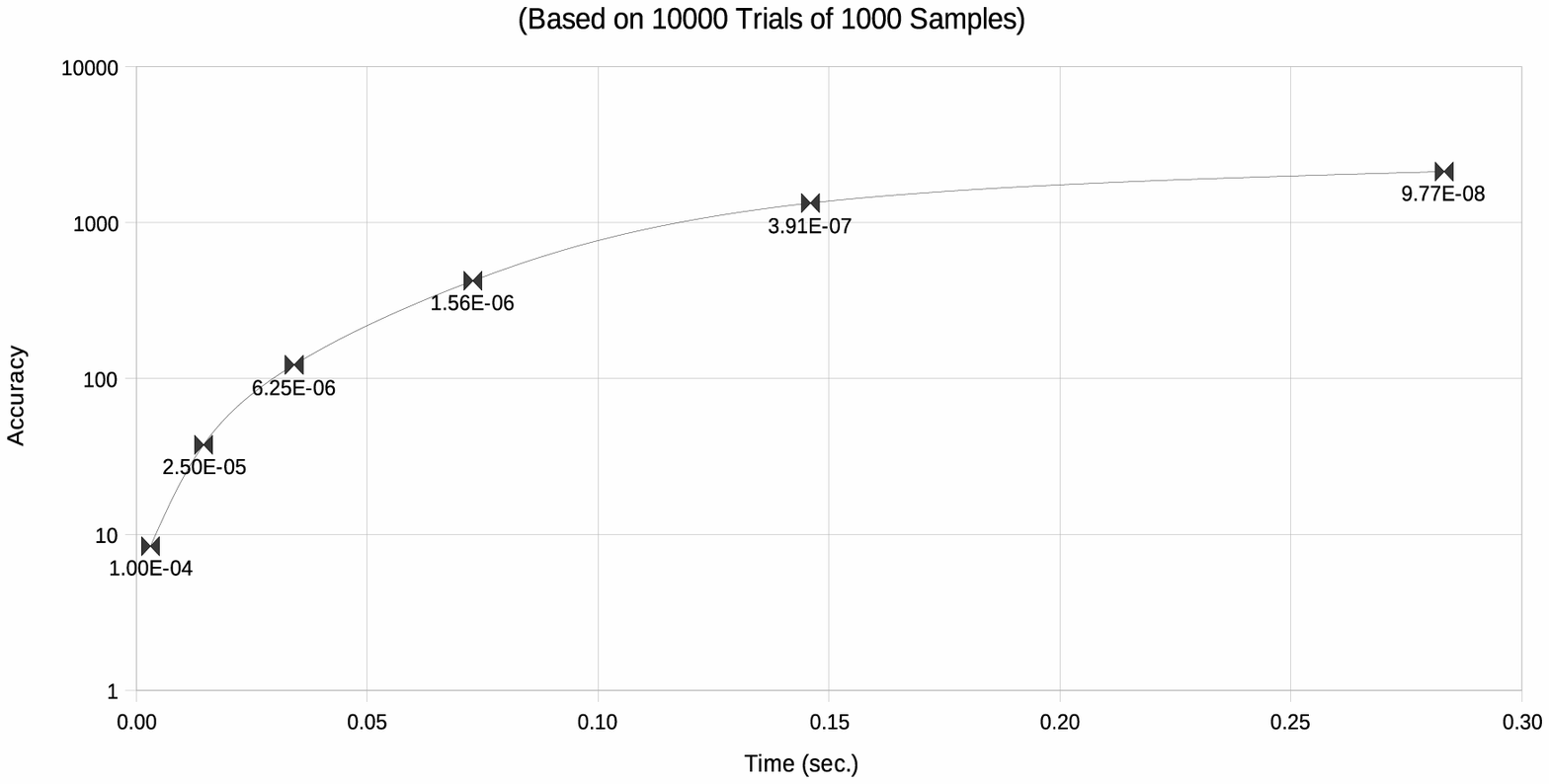}
\caption{Accuracy versus time trade-off in the ADAPT method (based on 10000 
 trials with 1000 samples). Node labels on graph are the tolerances logarithmic scale on vertical axis.}
\label{fig_accur_vs_time_adapt}
\end{minipage}
\end{figure}

Our next task is to compare the predictor-corrector method with our algorithm. 
For this we introduce the quantification of accuracy, which is 
defined as the inverse of the absolute relative bias. Furthermore, 
we saw from Figures \ref{fig_bias_pred_correct} and \ref{fig_bias_adapt} that 
using the predictor-corrector method with smaller interval size has the same 
directional effect as reducing the tolerance level in our algorithm, and both 
these actions result in increasing the time spent in simulation. Therefore the 
most ideal way of comparing the two methods is by plotting the accuracy, as 
defined above, versus the time spent in simulation. This is what is shown in 
Figure \ref{fig_pred_correct_vs_adapt}. 
As a byproduct of this analysis, we make a very interesting observation
regarding our method. It is exponentially rewarding 
in the initial part with a much steeper slope than the predictor-corrector 
method. Note that the accuracy is plotted on a logarithmic scale. We also note 
that both the methods taper off as we move to the right, with diminishing 
rewards. This may be attributed to the fact that reducing the bias substantially
below the simulation error is fruitless. 
This last observation brings us to the final figure of this section, 
Figure~\ref{fig_accur_vs_time_adapt}. 
In most risk management work, the computational budget associated with pricing a
financial derivative is limited. Based on that constraint, only a small number 
of simulation paths (typically between 1000--5000) are used for pricing. 
Of course, this results 
in a large value for the error associated with the Monte Carlo method. 
It is clearly pointless to control the bias to any order of 
magnitude below this error. 
Figure~\ref{fig_accur_vs_time_adapt} allows us to 
demonstrate that controlling the bias with a tolerance below 1.56e-06, say, 
has diminishing returns. 

\setlength{\tabcolsep}{1.75mm}
\begin{table} 
\caption{Number of intervals used in the calculation of the integral of $V$
         (tolerance = 0.000001).}
\label{table_num_intervals}
{\scriptsize
\begin{tabular}{c|cccccccccccc}\hline
right bin &	\multicolumn{12}{c}{Endpoint Variance}\\ \cline{2-13}	
boundaries	&	0.000001	&	0.0001	&	0.01	&	0.04	&	0.09	&	0.16	&	0.25	&	0.36	&	0.49	&	0.64	&	0.81	&	1	\\ \hline
8	&	0	&	0	&	0	&	0	&	0	&	0	&	0	&	0	&	0	&	0	&	0	&	0\\ 
16	&	2539	&	2507	&	1715	&	874	&	353	&	77	&	16	&	0	&	0	&	0	&	0	&	0\\ 
24	&	5098	&	5134	&	4978	&	4329	&	3295	&	1937	&	730	&	188	&	36	&	3	&	0	&	0\\ 
32	&	1886	&	1857	&	2490	&	3241	&	3732	&	3780	&	3008	&	1656	&	651	&	180	&	40	&	7\\ 
40	&	366	&	385	&	594	&	1016	&	1658	&	2437	&	3037	&	3027	&	2255	&	1364	&	608	&	156\\ 
48	&	96	&	97	&	183	&	382	&	637	&	1123	&	1887	&	2663	&	3032	&	2761	&	1986	&	1057\\ 
56	&	12	&	19	&	36	&	132	&	244	&	463	&	908	&	1517	&	2279	&	2801	&	2999	&	2453\\ 
64	&	3	&	0	&	4	&	22	&	70	&	141	&	298	&	646	&	1124	&	1744	&	2342	&	2820\\ 
72	&	0	&	1	&	0	&	2	&	8	&	33	&	85	&	203	&	419	&	726	&	1266	&	1973\\ 
80	&	0	&	0	&	0	&	2	&	1	&	6	&	21	&	59	&	139	&	293	&	536	&	1016\\ 
88	&	0	&	0	&	0	&	0	&	2	&	3	&	10	&	41	&	65	&	128	&	223	&	518\\ \hline
\end{tabular}
}
\end{table}

\begin{table}
\caption{Unused tolerance (0.000001) in the calculation of the integral of $V$.}
\label{table_unused_tol}
{\scriptsize
\begin{tabular}{c|cccccccccccc}\hline
right bin	&	\multicolumn{12}{c}{Endpoint Variance}\\ \cline{2-13}
boundaries	&	0.000001	&	0.0001	&	0.01	&	0.04	&	0.09	&	0.16	&	0.25	&	0.36	&	0.49	&	0.64	&	0.81	&	1	\\ \hline
8	&	0	&	0	&	0	&	0	&	0	&	0	&	0	&	0	&	0	&	0	&	0	&	0\\ 
16	&	2539	&	2507	&	1715	&	874	&	353	&	77	&	16	&	0	&	0	&	0	&	0	&	0\\ 
24	&	5098	&	5134	&	4978	&	4329	&	3295	&	1937	&	730	&	188	&	36	&	3	&	0	&	0\\ 
32	&	1886	&	1857	&	2490	&	3241	&	3732	&	3780	&	3008	&	1656	&	651	&	180	&	40	&	7\\ 
40	&	366	&	385	&	594	&	1016	&	1658	&	2437	&	3037	&	3027	&	2255	&	1364	&	608	&	156\\ 
48	&	96	&	97	&	183	&	382	&	637	&	1123	&	1887	&	2663	&	3032	&	2761	&	1986	&	1057\\ 
56	&	12	&	19	&	36	&	132	&	244	&	463	&	908	&	1517	&	2279	&	2801	&	2999	&	2453\\ 
64	&	3	&	0	&	4	&	22	&	70	&	141	&	298	&	646	&	1124	&	1744	&	2342	&	2820\\ 
72	&	0	&	1	&	0	&	2	&	8	&	33	&	85	&	203	&	419	&	726	&	1266	&	1973\\ 
80	&	0	&	0	&	0	&	2	&	1	&	6	&	21	&	59	&	139	&	293	&	536	&	1016\\ 
88	&	0	&	0	&	0	&	0	&	2	&	3	&	10	&	41	&	65	&	128	&	223	&	518\\ \hline
\end{tabular}
}
\end{table}

\clearpage

\clearpage

\clearpage

\section{Conclusion}\label{conclusion}
\setcounter{equation}{0}

In this paper we explored a variant of the methodology proposed by Broadie and 
Kaya in \cite{BroadieKaya}, to simulate the dynamics of the Heston model. 
As our method relies on the numerical computation of the integral of the 
variance process, it is subject to bias. The adaptive nature of our method
allows the efficient, practical control of this bias.

It is expected that for options on instruments with a large number of reset 
dates, the adaptive method of this paper should outperform the exact method of
Broadie and Kaya.

In summary, our method provides the following benefits:
\begin{enumerate}
\item Unlike finite-difference methods, our method cannot generate negative 
 values for the Bessel process and associated bridge.
\item Bias is efficiently controlled, as shown numerically by 
  Figures~\ref{fig_bias_adapt} and \ref{fig_accur_vs_time_adapt}.
\item It allows a much greater degree of flexibility than any of the other
  methods we have seen. This is advantageous where one can increase the
  tolerance level for middle-office risk-management work and reduce the
  tolerance level for front-office pricing. In other words, the tolerance level
  is a function of computational budget that one has at one's disposal.
\item Perhaps the most interesting feature of our method is a ``near-invariance"
  of the number of partitions (intermediate reset/coupon dates). Based on the
  observation that the most demanding part of our algorithm is the adaptive
  computation of the integral, we expect that the number of intermediate
  points we require for one big step of $T$ years is roughly equal to the
  total number of intermediate points required, had we decided to take $n$ steps
  of length $T/n$. In the latter case our method should perform much better than
  the  Broadie-Kaya method.
\item As our algorithm derives, in essence, from the Broadie-Kaya algorithm we
  can safely assume that all the extensions to jump diffusion models presented
  in \cite{BroadieKaya} should work with our algorithm as well. For the sake of
  brevity we do not reproduce the details in our paper.
\item Our method is straightforward to implement because many of the generators
  are now readily available in standard libraries.
\end{enumerate}

The most important extension that awaits investigation, is to higher dimensional
systems, for pricing options on equity baskets.

\section*{\Large Appendices}

\renewcommand{\theequation}{A.\arabic{equation}}
\renewcommand{\thesection}{A}
\newtheorem{thmA}{Theorem}[section]
\newtheorem{lemA}[thmA]{Lemma}
\newtheorem{defnA}[thmA]{Definition}
\newtheorem{propA}[thmA]{Proposition}
\newtheorem{corA}[thmA]{Corollary}
\setcounter{equation}{0}

\section*{Appendix A: Proof of Corollary \ref{cor_ianasif1}}
We will actually give the derivation in the opposite direction, from $X$ to $V$.
Let $\phi=\tau^{-1}$ so that $u=\tau(t)$ corresponds to $t=\phi(u)$; 
$u_0\equiv\tau(t_0)$. Also, for a constant $c>0$, to be determined,
denote $Y^c_t=e^{ct}X_{\tau(t)}\equiv e^{c\phi(u)}X_u$; so that
$Y^0_t=X_u$. Now, by (\ref{eqn_besselprocess}),
$$
X_u - X_{u_0} = \lambda(u-u_0) + 2\int_{u_0}^u\!\sqrt{X_s}\,dW_s
$$
so
\begin{eqnarray*}
X_{\tau(t)} - X_{\tau(t_0)} 
 &=& \lambda[\tau(t)-\tau(t_0)] 
       + 2\int_{\tau(t_0)}^{\tau(t)}\!
              \sqrt{X_{\tau\circ\phi(s)}}\,dW_{\tau\circ\phi(s)}\\
 &=& \lambda[\tau(t)-\tau(t_0)] 
       + 2\int_{t_0}^t\! \sqrt{X_{\tau(r)}}\,dW_{\tau(r)}
\end{eqnarray*}
where we have used a general time-substitution result for stochastic integrals
(see Proposition (30.10) in Chapter IV of \cite{RW})
to transform the integral in the second equality. Therefore
\begin{eqnarray}
dY^c_t 
 &=& cY^c_t\,dt 
     + e^{ct}[\lambda\,d\tau(t) + 2\sqrt{X_{\tau(t)}}\,dW_{\tau(t)}]\nonumber\\
 &=& cY^c_t\,dt + \lambda \frac{\sigma^2_V}{4}\,dt
     + 2e^{ct/2}\sqrt{{Y^c_t}}\,dW_{\tau(t)}.
\label{eqn_Y}
\end{eqnarray}

Let $Z$ be a given Brownian motion and take for $W$, the process defined by
$$
W_u = \frac{\sigma_V}{2}\int_{\phi(u_0)}^{\phi(u)}e^{-cs/2}\,dZ_s,
$$
so that 
\begin{equation}
W_{\tau(t)} = \frac{\sigma_V}{2}\int_{t_0}^t e^{-cs/2}\,dZ_s
\label{eqn_W}
\end{equation}
and, by It\^{o}'s formula,
$$
dW^2_{\tau(t)}
 =  \frac{\sigma^2_V}{4}
    \left[   
          \frac{4}{\sigma_V}W_{\tau(t)} e^{-ct/2}\,dZ_t + e^{-ct}\,dt.
    \right]
$$
Thus
$$
W^2_{\tau(t)} -  \frac{\sigma^2_V}{4}\int_{t_0}^t e^{-cs}\,dt
\equiv 
W^2_{\tau(t)} -  \frac{\sigma^2_V}{4}[\exp(-ct)-\exp(-ct_0)]/c
$$
is a martingale. With the choice $c=-\kappa$, we obtain that
$ W^2_{\tau(t)} -  \tau(t) $ is a martingale and changing variables back to 
$u$, that $W^2_u - u$ is a martingale (with respect to a different filtration, of
course).  Since $W$ itself is clearly a continuous martingale, we conclude from 
L\'{e}vy's theorem that $W$ is a Brownian motion on $[t_0, \infty)$.

Returning to (\ref{eqn_Y}) and substituting the differential form of 
(\ref{eqn_W}), we obtain, with $V\equiv Y^{-\kappa}$ and 
$\lambda=\frac{4\kappa\theta}{\sigma^2_V}$,
\begin{eqnarray*}
dV_t
 &=& -\kappa V_t\,dt + \kappa\theta \,dt
     + \sigma_V\sqrt{{V_t}}\,dZ_t.
\end{eqnarray*}
Finally, we identify $Z$ with $W^{(1)}$.  \endproof

\renewcommand{\theequation}{B.\arabic{equation}}
\renewcommand{\thesection}{B}
\newtheorem{thmB}{Theorem}[section]
\newtheorem{lemB}[thmB]{Lemma}
\newtheorem{defnB}[thmB]{Definition}
\newtheorem{propB}[thmB]{Proposition}
\newtheorem{corB}[thmB]{Corollary}
\setcounter{equation}{0}

\section*{Appendix B: Proof of Theorem \ref{thm_iantheorem2}}
Our starting point is the following representation taken from \cite{RY} (see 
Theorem 3.2 and its proof on pages 442--443, therein\footnote{Note that we have
replaced the arbitrary measure $\mu$ by $2\mu$ and corrected two typographic
errors on page 443: $\delta$ should be divided by 2 and the subscript 
$\rho^2(1)$ on $q$, should be $\sigma^2(1)$.}) for the Laplace functional of the
integral of a {\it BESQ} bridge, $X$, which starts at $x$ and ends at $y$:
$$
 \E_{xy}\!\left[\exp\left\{-\int_0^1\!X(u)\,d\mu(u)\right\}\right]
 \equiv 
 \E_x\!\left[\exp\left\{-\int_0^1\!X(u)\,d\mu(u)\right\}\,\Big|\,
             X(1)=y
       \right], \quad x,y\geq 0
$$
where $\mu$ is a Radon measure on $[0,\infty)$ with support in $[0,1]$.
\begin{thmB} Let $\mu$ be a Radon measure on $[0,\infty)$ with support in 
$[0,1]$, and set
$$
 F(x,y,\mu):=
  \E_{xy}\!\left[\exp\left\{-\int_0^1\!X(u)\,d\mu(u)\right\}\right].
$$
Then
\begin{eqnarray*}
 \lefteqn{F(x,y,\mu)  
 =
     \left[\phi(1)\int_0^1\!\phi(u)^{-2}du\right]^{-1}\cdot
     \frac{ 
       I_{\nu}\!\left(\frac{\sqrt{xy}}{\phi(1)\int_0^1\!\phi(u)^{-2}du}\right) 
     }
     { I_{\nu}\!\left(\sqrt{xy}\right) } 
 } \\ [2mm]
 &&\ \cdot
   \exp\!\left\{
            \frac{x}{2}
            \left[\phi'(0)-\left(\int_0^1\!\phi(u)^{-2}du\right)^{-1}+1\right]
          \right\}
   \cdot
   \exp\!\left\{
            \frac{y}{2}
            \left[1-\left(\phi(1)^2\int_0^1\!\phi(u)^{-2}du\right)^{-1}\right]
          \right\} 
\end{eqnarray*}
where $I_{\nu}$ is the modified Bessel function of order $\nu$ and $\phi$ is the
unique solution (in the sense of generalised functions) of the ODE
$$
 \phi''=2\mu\cdot\phi,\ \phi(0)=1,\ \phi\geq 0,\ \phi\ 
    \mbox{nonincreasing on $[0,\infty)$}.
$$
Consequently, $\phi$ is convex and right-differentiable with a 
right-continuous, nonpositive right-derivative.
\end{thmB}

This result can be transferred to general interval $[0,\tau]$, by using the 
following result (see \cite{PY} or \cite{RY}): If $Q^{\tau}_{x,y}$ denotes the 
law of $X$ under which it is a Bessel bridge, starting at $x$ and ending at $y$
(at time $\tau$), then $Q^{\tau}_{x,y}$ is also the $Q^1_{x/\tau,y/\tau}$-law of
$\tau X(u/\tau)$, on $[0,\tau]$. The only case of interest to us, is when
$\mu$ has a density with respect to Lebesgue measure, necessarily of the form,
$m(u)\Bid_{[0,\tau]}(u)$, $0\leq u <\infty$. With a slight abuse of 
notation, we write $F(x,y,m)$ instead of $F(x,y,\mu)$ in this setting. The 
details of the transformation are as follows. Given the density $m$, we set 
$m_{\tau}(u):=\tau^2m(\tau u)$, $0\leq u \leq 1$; $m_{\tau}(u):=0$, for 
$u>1$. Using the cited equivalence of laws and then making the change of 
variables, $u\mapsto \tau u$, yields (with $\E_{xy}^{\tau}$ denoting the 
$Q^{\tau}_{x,y}$ expectation):
\begin{eqnarray*} 
 \E_{xy}^{\tau}\left[\exp\left\{-\int_0^{\tau}\!X(u)m(u)\,du\right\}\right]
 &=& \E_{x/\tau,y/\tau}^1
     \left[\exp\left\{-\int_0^{\tau}\!\tau X(u/\tau)m(u)\,du\right\}\right] 
     \\
 &=& \E_{x/\tau,y/\tau}^1
     \left[\exp\left\{-\int_0^1\!X(u)m_{\tau}(u)\,du\right\}
     \right] \\
 &=& F\left(\frac{x}{\tau},\frac{y}{\tau},m_{\tau}\right).
\end{eqnarray*} 

There is one more technical result which is needed to completely localise the
problem to the support of $\mu$. Again we restrict attention to the case where
$\mu$ has a density $m$ which we assume is continuous on its support, the 
interval $[0,1]$. In that case, the ODE is
$$
 \phi''= 2m\Bid_{[0,1]}\cdot\phi,\ \phi(0)=1,\ \phi\geq 0,\ 
           \phi\ \mbox{nonincreasing.}
$$
Standard regularity theory yields that $\phi$ is smooth ($C^2$) on $[0, 1)$ and
satisfies the equation, $\phi''=2m\phi$ in the classical sense thereon. 
Of course, $\phi$ is constant on $(1,\infty)$ and being continuous everywhere, 
the constant value is $\phi(1)$. We now show that $\phi$ satisfies a Neumann 
boundary condition at $u=1$.
\begin{lemB} The left-hand derivative, $\phi'(1^-)=0$, so that $\phi$ is $C^1$ 
smooth across $u=1$.
\end{lemB}

{\bf Proof\ } Let $g\in C^{\infty}_0((0,\infty))$ be a test function. The ODE 
(aside from the boundary and side conditions) means that
$$
\frac{1}{2}\int_0^{\infty}\!\phi(u)g''(u)\,du 
 = \int_0^{\infty}\!m(u)\Bid_{[0,1]}(u)\phi(u)g(u)\,du
 = \int_0^1\!m(u)\phi(u)g(u)\,du.
$$
Also,
\begin{eqnarray*}
 \frac{1}{2}\int_0^{\infty}\!\phi(u)g''(u)\,du 
 &=& \int_0^1\!\phi(u)g''(u)\,du + \int_1^{\infty}\!\phi(u)g''(u)\,du\\ 
 &=& \int_0^1\!\phi(u)g''(u)\,du + \phi(1)\int_1^{\infty}\!g''(u)\,du\\
 &=& \int_0^1\!\phi(u)g''(u)\,du - \phi(1)g'(1)\\
 &=& \phi(1)g'(1) - \int_0^1\!\phi'(u)g'(u)\,du - \phi(1)g'(1)\\
 &=& -\phi'(1^-)g(1) + \int_0^1\!\phi''(u)g(u)\,du\\ 
 &=& -\phi'(1^-)g(1) + \int_0^1\!m(u)\phi(u)g(u)\,du.
\end{eqnarray*}
Therefore $\phi'(1^-)g(1)=0$ for all $g\in C^{\infty}_0((0,\infty))$, which 
implies that $\phi'(1^-)=0$. \endproof

Thus we arrive at the final formulation of the required Laplace transform:
\begin{corB}\label{cor_Laplace} 
Let $m_{\tau}=a[b+cu]^{-2}$, $0\leq u\leq 1$, where
$$
 a = \frac{8\theta \tau^2}{\sigma_V^2},\ b= 1 + \frac{4\kappa\tau_L}{\sigma_V^2},\ 
 c = \frac{4\kappa\tau}{\sigma_V^2}
$$
($m_{\tau}=2\theta\tau^2 w(\tau u)$, where $w$ was introduced at 
(\ref{eqn_weight_canonical})) and set
$$
 L(\theta):=
  \E_{xy}\!\left[\exp\left\{-\theta\int_0^{\tau}\!X(u)w(u)\,du\right\}\right].
$$
Then
\begin{eqnarray}
 L(\theta)
 &=&
   \left[\phi(1)\int_0^1\!\phi(u)^{-2}du\right]^{-1}\cdot
   \frac{
     I_{\nu}\!\left(\frac{\sqrt{xy}}{\tau\phi(1)\int_0^1\!\phi(u)^{-2}du}\right)
   }
   { I_{\nu}\!\left(\frac{\sqrt{xy}}{\tau}\right) } 
 \nonumber\\ [2mm]
 & &\ \cdot\,
 \exp\!\left\{
            \frac{x}{2\tau}
            \left[\phi'(0)-\left(\int_0^1\!\phi(u)^{-2}du\right)^{-1}+1\right]
       \right\} \nonumber \\ [2mm]
 & &\ \cdot\,
    \exp\!\left\{
            \frac{y}{2\tau}
            \left[1-\left(\phi(1)^2\int_0^1\!\phi(u)^{-2}du\right)^{-1}\right]
          \right\} \nonumber \\ [1mm]
\label{eqn_Lapl}
\end{eqnarray}
where $I_{\nu}$ is the modified Bessel function of order $\nu$ and $\phi$ is the
unique solution to the BVP:
\begin{equation}
 \phi''=m_{\tau}(u)\cdot\phi,\ \phi(0)=1,\ \phi'(1)=0.
\label{eqn_ode_phi}
\end{equation}
\end{corB}

For the first two moments, we are interested in the coefficients of
$-\theta$ and $\theta^2/2$ in the Taylor expansion of $L(\theta)$ about $\theta=0$, since 
$e^{-r}=1-r+r^2/2+\cdots$ and thus the left-hand side of (\ref{eqn_Lapl}) equals
\begin{eqnarray*}
1 - \theta\cdot\E_{xy}\!\left[ \int_0^\tau\!X(u)w(u)\,du\right]
  +\frac{\theta^2}{2}\cdot
  \E_{xy}\!\left[ \left(\int_0^{\tau}\!X(u)w(u)\,du\right)^2\right]
  + \cO(\theta^3).
\end{eqnarray*}
Accordingly, we work out the Taylor expansion of the right-hand side of
(\ref{eqn_Lapl}) to second order.

With reference to (\ref{eqn_ode_phi}), the function $\phi$ depends implicitly on
$\theta$ through the constant, $a$, which appears in the definition of the function
$m_{\tau}$. We make this dependence explicit in our notation, by writing 
$\phi(u;\theta)$.
Clearly $\phi(\cdot\,;0)\equiv 1$; so $\phi'(0;0)=0$. (Differentiation with 
respect to $u$ will continue to be denoted by a prime ($'$) superscript; 
differentiation with respect to $\theta$ will be written explicitly; e.g., as a 
partial derivative, $\partial/\partial\theta$.) Thus we set
\begin{eqnarray*}
 \phi(1;\theta) &=& 1 + A_1\theta + A_2\theta^2 + \cO(\theta^3)\\
 \int_0^1\!\phi(u;\theta)^{-2}du &=& 1 + B_1\theta + B_2\theta^2 + \cO(\theta^3)\\
 \phi'(0;\theta) &=& C_1\theta + C_2\theta^2 + \cO(\theta^3)
\end{eqnarray*}
leaving the determination of the coefficients, $A_i$, $B_i$, $C_i$ (i=1,2) for
later.

Expansion of the right-hand side of (\ref{eqn_Lapl}) in powers of $\theta$, can be
effected in a few stages. The terms involving a multiplicative inverse, like
$(\int_0^1\!\phi(u)^{-2}du)^{-1}$, can be expanded using the expansion for 
$\phi$ and the geometric series expansion, 
$r^{-1}=(1-[1-r])^{-1}=1+[1-r]+[1-r]^2 + \cO([1-r]^3)$, for $r$ close to 1. 
The two exponentials can be combined and then 
handled with the usual expansion, $e^r=1+r+r^2+\cO(r^3)$, for $r$ close to 0.
The ratio of modified Bessel functions can be handled, using the following two 
identities for modified Bessel functions (see 9.6.1, 9.6.26 in \cite{AS}):
\begin{eqnarray}
I''_{\nu}(r)
 &=& \left(1+\frac{\nu^2}{r^2}\right)I_{\nu}(r) - \frac{1}{r} I'_{\nu}(r)
     \label{eqn_I_id1}\\
I'_{\nu}(r)&=& I_{\nu+1}(r) + \frac{\nu}{r}I_{\nu}(r)\label{eqn_I_id2}.
\end{eqnarray}
We can substitute (\ref{eqn_I_id2}) into (\ref{eqn_I_id1}) and divide
both identities by $I_{\nu}(r)$ to obtain the following ones in terms of
the so-called {\it Bessel quotient} function,
$R_{\nu} = I_{\nu+1}(r)/I_{\nu}(r)$:
\begin{eqnarray*}
\frac{I''_{\nu}(r)}{I_{\nu}(r)}
 &=& 1+\frac{\nu^2-\nu}{r^2} - \frac{1}{r} R_{\nu}(r) \\
\frac{I'_{\nu}(r)}{I_{\nu}(r)}
 &=& \frac{\nu}{r} + R_{\nu}(r).
\end{eqnarray*}
We can apply these identities to (\ref{eqn_Lapl}) by writing 
$r_0=\sqrt{xy}/\tau$ and 
$r=r_0/\phi(1)\int_0^1\!\phi(u)^{-2}du$, and expressing the ratio of Bessel 
functions in (\ref{eqn_Lapl}), in the form
$$
 \frac{I_{\nu}(r)}{I_{\nu}(r_0)} = \frac{I_{\nu}(r_0+[r-r_0])}{I_{\nu}(r_0)}
 = 
 \frac{
  I_{\nu}(r_0) + I'_{\nu}(r_0)[r-r_0]) + I''_{\nu}(r_0)[r-r_0]^2/2 
    + \cO\!\left([r-r_0]^3\right)
 }
 {I_{\nu}(r_0)}
$$
and noting that
$$
 r-r_0 = r_0[(\phi(1)\int_0^1\!\phi(u)^{-2}du)^{-1} - 1]
$$
has an expansion in $\theta$, without constant term. 

The remaining details are straightforward but tedious algebra which is omitted.
The final result is the expression in Theorem \ref{thm_iantheorem2}.

We now turn to the calculation of the constants, $A_i$, $B_i$, $C_i$ (i=1,2), in
terms of the constants $a,b,c$. Denoting
\begin{eqnarray*}
 \Delta_{\theta}(u) 
   &:=& \frac{ \partial\phi(u;\theta) }{ \partial\theta },\quad  
        \Delta(u)\ :=\ \Delta_0(u)\\
 \Gamma_{\theta}(u) 
   &:=& \frac{ \partial^2\phi(u;\theta) }{ \partial\theta^2 },\quad  
        \Gamma(u)\ :=\ \Gamma_0(u),
\end{eqnarray*}
we then have
\begin{equation}
 A_1=\Delta(1),\ A_2=\frac{1}{2}\Gamma(1);\ 
 C_1=\Delta'(0),\ C_2=\frac{1}{2}\Gamma'(0).
\label{eqn_AC_prelim}
\end{equation}
Also, for $B_1,B_2$, note that
\begin{eqnarray*}
 \frac{d}{d\theta} \int_0^1\!\phi(u;\theta)^{-2}du 
  &=&\int_0^1-2\phi(u;\theta)^{-3}\,\frac{ \partial\phi(u;\theta) }{ \partial\theta }\,du\\
 \frac{d^2}{d\theta^2} \int_0^1\!\phi(u;\theta)^{-2}du
  &=& \int_0^1 
        6\phi(u;\theta)^{-4}\!
          \left[\frac{ \partial\phi(u;\theta) }{ \partial\theta }\,\right]^2
        -2\phi(u;\theta)^{-3}\,\frac{ \partial^2\phi(u;\theta) }{ \partial\theta^2 }
       du.\\
\end{eqnarray*}
Evaluating these results at $\theta=0^+$, we obtain
\begin{eqnarray}
 B_1 
  &=& \left.\frac{d}{d\theta}\right|_{\theta=0^+} \int_0^1\!\phi(u;\theta)^{-2}du
         \ =\ -2 \int_0^1\!\Delta(u)\,du, \label{eqn_B1_prelim}\\
 B_2 
  &=& \left.\frac{1}{2}\frac{d^2}{d\theta^2}\right|_{\theta=0^+} 
      \int_0^1\!\phi(u;\theta)^{-2}du  
  \ =\ \int_0^1\!3\Delta(u)^2 - \Gamma(u)\,du. \label{eqn_B2_prelim}
\end{eqnarray}
The functions $\Delta$ and $\Gamma$ can be found by differentiating 
($\frac{\partial}{\partial\theta}$) the ODE and boundary conditions for $\phi$ and 
then solving the resulting, very simple problems at $\theta=0^+$. To that end, we
bring out the $\theta$-dependence of $m_{\tau}$ explicitly by writing
$$
 m_{\tau}\equiv \theta M_{\tau}
$$
with $M_{\tau}$ independent of the parameter, $\theta$. Then,
\begin{eqnarray*}
 &&\Delta''_{\theta} = \frac{\partial}{\partial\theta}\phi'' 
 = \frac{\partial}{\partial\theta}(\theta M_{\tau}\phi) 
 = M_{\tau}\phi + \theta M_{\tau}\frac{\partial\phi}{\partial\theta}
 \Longrightarrow
 \Delta''=M_{\tau};\mbox{\ also\ }\Delta(0)=0,\ \Delta'(1)=0;\\
 &&\Gamma''_{\theta} = \frac{\partial^2}{\partial\theta^2}\phi''
 = 2M_{\tau}\frac{\partial\phi}{\partial\theta} 
    + \theta M_{\tau}\frac{\partial^2\phi''}{\partial\theta^2}
 \Longrightarrow
 \Gamma'' = 2M_{\tau}\Delta;\mbox{\ also\ }\Gamma(0)=0,\ \Gamma'(1)=0.
\end{eqnarray*}
Integrating and using the boundary conditions, we obtain:
\begin{eqnarray}
 \Delta'(u) 
   &=& -\int_u^1\!M_{\tau}(v)\,dv,\quad
 \Delta(u)
   \ =\ -\int_0^u\!\Delta'(v)\,dv; \label{eqn_delta_ints}\\
 \Gamma'(u) 
   &=& -2\int_u^1\!M_{\tau}(v)\Delta(v)\,dv,\quad
 \Gamma(u) 
   \ =\ \int_0^u\! \Gamma'(v)\,dv. \label{eqn_gamma_ints}
\end{eqnarray}
\begin{propB} \label{prop_integralsABC}
With $a,b,c$ as in Corollary \ref{cor_Laplace}, 
$A:=a/\theta$, and $\log^2$ denoting the square of the $\log$ function,
\begin{eqnarray}
 A_1 
  &=& \frac{A}{c}\left[\frac{1}{b+c} + \frac{1}{c}\log[b/(b+c)]\right]\label{eqn_A1}\\
 A_2 
  &=& \frac{A^2}{c^4(b+c)}
      \left[
        (b+c)\log^2[b/(b+c)] - 3b\log[b/(b+c)] - \frac{c(3b+2c)}{b+c}
      \right]
     \label{eqn_A2}\\
 B_1
  &=& -\frac{2A}{c^2}
       \left[\frac{2b+3c}{2(b+c)} + \frac{b+c}{c}\log[b/(b+c)]\right]
        \label{eqn_B1}\\
 B_2
  &=& \frac{-A^2}{c^5(b+c)} \nonumber \\
  & & \quad\cdot
      \left[
            2(b+c)^2\log^2[b/(b+c)] 
            - 2(3b^2 + 8cb + 4c^2)\log[b/(b+c)]
            - c(6b + 11c)
      \right] 
 \nonumber \\
 && \label{eqn_B2}\\
 C_1
  &=& -\frac{A}{b(b+c)} \label{eqn_C1}\\
 C_2
  &=& \frac{A^2}{c^3(b+c)}\left[\frac{c(2b+c)}{b(b+c)} + 2\log[b/(b+c)]\right].
      \label{eqn_C2}
\end{eqnarray}
\end{propB}
{\bf Proof\ } 
All of the integrals in (\ref{eqn_delta_ints}) and (\ref{eqn_gamma_ints}) are 
straightforward to evaluate, as well as the integrals (\ref{eqn_B1_prelim}) and
(\ref{eqn_B2_prelim}),
for $B_1$ and $B_2$. We omit the elementary calculus and just state the end
results, as they can be easily verified by differentiation and checking a 
boundary condition. Then one can use (\ref{eqn_AC_prelim}) to obtain $A_1, A_2,
C_1$ and $C_2$.
\begin{eqnarray*}
 \Delta'(u)
   &=&  \frac{A}{c}\left[(b+c)^{-1} - (b+cu)^{-1}\right]\\
 \Delta(u)
   &=& \frac{A}{c}\left[(b+c)^{-1}u +\frac{1}{c}\log[b/(b+cu)]\right]\\
 \int_0^u\!\!\Delta(v)\,dv
    &=& \frac{A}{c}
        \left[
            \frac{u^2}{2(b+c)} + \frac{(1+\log b)u}{c} -\frac{b+cu}{c^2}\log[b+cu]
            + \frac{b}{c^2}\log b
        \right] \\
 \Gamma'(u)
   \!&=&\! \frac{2A^2}{c^3(b+c)} 
        \left[
           \log\!\!\left[\frac{b}{b+c}\right]-\log[b+c] - \frac{2b+c}{b+c} 
           + \frac{2b+c}{b+cu} 
       \right.\\
   & &\qquad\qquad\quad
        \left.
           - \frac{b+c}{b+cu}\log\!\!\left[\frac{b}{b+cu}\right] 
           + \log[b+cu]
        \right]  \\
 \Gamma(u)
   \!\!&=&\!\!\frac{2A^2}{c^3(b+c)}
       \left[
           \frac{b+c}{2c}(\log b)^2 -\frac{3b+c}{b+c}\log b 
           + \frac{2b+c -(b+c)\log b}{c}\log[b+cu] 
       \right.\\
   & &\qquad\qquad\, 
       \left.
           +\ \frac{b+c}{2c}(\log[b+cu])^2
           + \left(\log\!\frac{b}{(b+c)^2} - \frac{3b+2c}{b+c}\right)u
       \right.\\
   & &\qquad\qquad\, 
       \left.
           + \frac{b+cu}{c}\log[b+cu]
       \right]\!;
\end{eqnarray*}
\begin{eqnarray*}
 \lefteqn{\int_0^u\!3\Delta(v)^2-\Gamma(v)\,dv}\\
   &=& \frac{A^2 u^3}{c^2(b+c)^2}
       + \frac{A^2}{c^5(b+c)^2}
         \left[
               2c^3 + 3bc^2-(b+c)c^2\log b + 2(b+c)c^2\log[b+c] 
         \right]\!u^2 \\
   & &\quad
       +\ \frac{2A^2b(3b + 2c)\log b}{c^4(b+c)^2}\,u
       + \frac{A^2\left[-4(b+c)b^2 + (5b+3c)b^2\log b - 6bc(b+c)\right]}
            {c^5(b+c)^2} \\
   & &\quad
       +\ \frac{A^2}{c^5(b+c)}\, (b+cu)^2
         \big[ 3\log[b/b+cu] - \log[b+cu] + 2 \big] \\
   & &\quad
       +\ \frac{A^2}{c^5(b+c)^2}\,(b+cu)
          \big[
               2(b+c)^2\log^2[b/b+cu]
              -2\left(b^2 - 3cb - 3c^2\right)\log[b/b+cu] 
          \\
  & &\qquad\qquad\qquad\qquad\qquad\qquad\qquad\qquad
              -\ 2b(2b+c)\log[b+cu]
              +2(b+3c)(b+c) 
          \big]. 
\end{eqnarray*}
\endproof

\renewcommand{\theequation}{C.\arabic{equation}}
\renewcommand{\thesection}{C}
\newtheorem{thmC}{Theorem}[section]
\newtheorem{lemC}[thmC]{Lemma}
\newtheorem{defnC}[thmC]{Definition}
\newtheorem{propC}[thmC]{Proposition}
\newtheorem{corC}[thmC]{Corollary}
\setcounter{equation}{0}

\section*{Appendix C: Explicit Laplace transform}
In the previous appendix, we avoided solving the BVP (\ref{eqn_ode_phi}) for 
$\phi$ and then calculating $\int_0^1\!\phi(u)^{-2}\,du$. In order to extricate 
the moments of $\int_0^{\tau}X(u)w(u)\,du$ from its Laplace transform, it was 
sufficient to apply the method of variation of parameters, which led to 
expressions for the moments in terms of $\phi$ and its derivatives with respect 
to the Laplace parameter, $\theta$, at $\theta=0$. This method could work for 
more general weight functions than our specific $w$.

However, to distributionally validate the {\it ADAPT} schemes, it is most 
convenient to have an explicit form for the Laplace transform (\ref{eqn_Lapl}), 
and for that we now solve the BVP.

\begin{propC}\label{prop_BVPsoln}
The solution to the BVP 
\begin{equation}
 \phi''=m_{\tau}(u)\cdot\phi,\ \phi(0)=1,\ \phi'(1)=0
\label{eqn_ode_phi_again}
\end{equation}
is
\begin{eqnarray}
\phi(u) &=& \epsilon b^p(b+cu)^{-p} + (1-\epsilon) b^q(b+cu)^{-q} 
          \label{eqn_phi_soln}\\
p &=& \frac{-1+\sqrt{c^2+4a}/c}{2},\ q \ =\ \frac{-1-\sqrt{c^2+4a}/c}{2} 
        \label{eqn_powers}\\
\epsilon &=& -qb^q/[pb^p(b+c)^{-\frac{\sqrt{c^2+4a}}{c}} - qb^q].
            \label{eqn_eps}
\end{eqnarray}
where $a,b,c$ were defined in Corollary \ref{cor_Laplace}.
\end{propC}
{\bf Proof\ } It is a simple matter to verify that the function described by
(\ref{eqn_phi_soln})--(\ref{eqn_eps}) satisfies the ODE and boundary conditions
(\ref{eqn_ode_phi_again}). A sketch of the method of solution is as follows.
We set $\psi(u)=\phi(u/\sqrt{a})$ and then $\psi(u)=f(r)$ where 
$r=[b+\delta u]^{-1}$ and $\delta=c/\sqrt{a}$. This leads to the following BVP 
for $f$ on the interval $[r_e, b^{-1}]$, $r_e\equiv [b+c]^{-1}$:
$$
 r^2\frac{d^2f(r)}{dr^2} + 2r\frac{df(r)}{dr} = \delta^{-2}f(r),\quad 
  f(b^{-1})=1,\ f'(r_e)=0.
$$ 
Seeking a solution of the ODE alone, in the form $r^p$, leads to the condition
$$
 p^2 + p -\delta^{-2} = 0
$$
which has the two solutions (\ref{eqn_powers}). The function
$$
 f(r) = \epsilon b^p r^p + (1-\epsilon)b^q r^q
$$
then satisfies the boundary condition, $f(b^{-1})=1$, for any $\epsilon$, while
the choice, (\ref{eqn_eps}), guarantees that the other boundary condition,
$f'(r_e)=0$, is satisfied. Unwinding the definitions from $f$ back to $\phi$ 
yields (\ref{eqn_phi_soln}). \endproof

\begin{corC}\label{cor_phi_stuff}
With $p,q$ defined at (\ref{eqn_powers}),
\begin{eqnarray}
 \phi'(0) 
  &=& -\frac{a}{bc}\,
      \frac{b^q - b^p(b+c)^{-\sqrt{c^2+4a}/c}}
         {p b^p(b+c)^{-\sqrt{c^2+4a}/c} - q b^q} \label{eqn_phi0} \\
 \phi(1)
  &=& \frac{1}{b}\,
      \frac{p (b+c)^{-p} - q(b+c)^{-q}}
         {p b^p (b+c)^{-\sqrt{c^2+4a}/c} - qb^q} \label{eqn_phi1} \\
 \int_0^1\!\!\phi(u)^{-2}\,du
  &=& \frac{(b(b+c))^{q-p}c^2}
         {8\left(c^2 + 4a\theta\right)^{3/2}
           \left(
                  \left(c^2 + 2a\theta\right)(b(b+c))^{p-q}
                  + a\left(b^{2(p-q)} + (b+c)^{2(p-q)}\right) \theta
           \right)
         }\nonumber \\
  & & \quad\cdot
         \left(
                (p-q)b^{p-q+\frac{1}{2}}
                - b^{p-q+\frac{1}{2}} + (b+c)^{p-q}\sqrt{b}
                + (b+c)^{p-q}\sqrt{b}(p-q)
         \right)^2 
         \nonumber \\
  & & \quad\cdot
         \left(
                2c^2(p-q)(b(b+c))^{p-q}
                + (b+c)^{2(p-q)}\left(c^2 + 4a\theta\right)
         \right.  \nonumber \\
  & & \qquad\ 
          \left.
                 -\ b^{2 (p-q)}
                  \left(
                         4a\theta + c\left(c + c(p-q)\right)
                  \right)
                - c(b+c)^{2(p-q)}c(p-q)
         \right)
  \label{eqn_int_phi}
\end{eqnarray}
\end{corC}
{\bf Proof\ } The results (\ref{eqn_phi0}) and (\ref{eqn_phi1}) are 
straightforward calculations. On the other hand, the derivation of 
(\ref{eqn_int_phi}) is not straightforward; it was effected with the aid of 
\textit{Mathematica}$^{\textsuperscript{\textregistered}}$.
 \endproof

%\algoblankpage

\begin{thebibliography}{9}
\bibitem{AS} Abramowitz, M., Stegun, I. (Eds.): Handbook of Mathematical 
  Functions. Dover Publications, New York (1970) 
\bibitem{AhrensDieter} Ahrens, J. H., Dieter, U.: Computer generation of   Poisson deviates from modified normal distributions. ACM Trans. Math. 
  Software, 8, No. 2, 163--179 (1982)
\bibitem{Andersen} Andersen, L.: Efficient simulation of the Heston 
  stochastic volatility model. Journal of Computational Finance, Vol. 11, No. 3,
  1--42 (2008)
\bibitem{Borodin} Borodin, A., Salminen, P.: Handbook of Brownian Motion -- Facts
  and Formulae. Birkh\:{a}user, Basel (1996)  
\bibitem{BroadieKaya} Broadie, M., Kaya, \:{O}.: Exact simulation of 
  stochastic volatility and other affine jump processes. Operations Research,
  Vol. 54, Issue 2, 217--231 (2006) 
\bibitem{Hormann} H\"ormann, W.: The transformed rejection method for
  generating Poisson random variables. Insurance, Mathematics and Economics, 12,
  39--45 (1993)  
\bibitem{KempKemp} Kemp, C. D., Kemp, A. W.: Poisson random variate 
  generation. Appl. Statist., 40, No. 1, 143--158 (1991)
\bibitem{Roman} Campolieti,G. Makarov, R.: Pricing path-dependent options 
  on state-dependent volatility models with a Bessel bridge. IJTAF Vol. 10, 
  Issue 01, 51--88 (2007)
\bibitem{CIR} Cox, J.C., Ingersoll, J.E., Ross, S.A.: A theory of the 
  term structure of interest rates. Econometrica, Vol. 53, No. 2, 385--407 (1985)
\bibitem{Gatheral} Gatheral, J.: The Volatility Surface: A Practitioner's Guide.
  John Wiley \& Sons, Hoboken N.J. (2006)
\bibitem{Heston} Heston, S.: A closed-form solution of options on assets 
  with stochastic volatility with applications to bond and currency options. The  Review of Financial Studies, Vol. 6, No. 2, 327--343 (1993)
\bibitem{PY} Pitman, J., Yor, M.: A decomposition of Bessel bridges.
  Z. W. verw. Gebiete 59, 425--457 (1982)
\bibitem{RY} Revuz, D., Yor, M.: Continuous Martingales and Brownian Motion.
  (2nd edition) Springer-Verlag (1994)
\bibitem{RW} Rogers, L.C.G., Williams, D.:  Diffusions, Markov Processes and 
  Martingales: Volume 2, It\^{o} Calculus (2nd edition).
  Cambridge University Press (2000) 
\bibitem{Soni} Soni, R. P.: On an inequality for modified Bessel 
  functions. J. Math. Phys. 44, 406--407 (1965) 
\bibitem{Yuan} Yuan, L. Kalbfleisch, J.: On the Bessel distribution and 
  related problems. Ann. Inst. Statist. Math. Vol. 52, No. 3, 438--447 (2000)
\end{thebibliography}
\end{document}